\documentclass[useAMS,usegraphicx]{mn2e}
\usepackage{lscape}
\usepackage{latexsym}
\usepackage{epsfig}
\begin{document}

\title[Correlations in a complete sample of Swift GRBs]{A complete sample of bright Swift Gamma--Ray Bursts: X--ray afterglow luminosity and its correlation with the prompt emission}

\author[P. D'Avanzo et al.]{P. D'Avanzo$^{1}$\thanks{E-mail: paolo.davanzo@brera.inaf.it}, R. Salvaterra$^{2}$, B. Sbarufatti$^{1}$, L. Nava$^{3}$, A. Melandri$^{1}$, \newauthor M. G. Bernardini$^{1}$, S. Campana$^{1}$, S. Covino$^{1}$, D. Fugazza$^{1}$, G. Ghirlanda$^{1}$, \newauthor G. Ghisellini$^{1}$, V. La Parola$^{4}$, M. Perri$^{5}$, S. D. Vergani$^{1}$ \& G. Tagliaferri$^{1}$\\
$^{1}$INAF - Osservatorio Astronomico di Brera, via Emilio Bianchi 46, I-23807, Merate (LC), Italy\\
$^{2}$INAF - IASF Milano, via E. Bassini 15, I-20133, Milano, Italy\\
$^{3}$APC, Univ Paris Diderot, CNRS/IN2P3, CEA/Irfu, Obs de Paris, Sorbonne Paris Cit\'e, France\\
$^{4}$INAF - IASF Palermo, via Ugo La Malfa 153, I-90146, Palermo, Italy\\
$^{5}$ASI - Science Data Centre, via Galileo Galilei, I-00044, Frascati, Italy}

\def\simg{\mathrel{%
      \rlap{\raise 0.511ex \hbox{$>$}}{\lower 0.511ex \hbox{$\sim$}}}}
\def\siml{\mathrel{%
      \rlap{\raise 0.511ex \hbox{$<$}}{\lower 0.511ex \hbox{$\sim$}}}}
\def\Mesz{M\'esz\'aros~}
\def\ie{i.e$.$~} \def\eg{e.g$.$~} \def\etal{et al$.$~} 
\def\eq{eq$.$~} \def\eqs{eqs$.$~} \def\deg{^{\rm o}} \def\dd{{\rm d}}
\def\beq{\begin{equation}} \def\eeq{\end{equation}}
\def\epsel{\varepsilon_e} \def\epse1{\varepsilon_{e,-1}} 
\def\epsmag{\varepsilon_B} \def\eBzero{\tilde{\varepsilon_B}} 
\def\eBone{\varepsilon_{B,-1}} \def\eBtwo{\varepsilon_{B,-2}} 
\def\eBthree{\varepsilon_{B,-3}} \def\eBfour{\varepsilon_{B,-4}}
\def\E53{E_{53}} \def\nex0{n_{*,0}} \def\Gm02{\Gamma_{0,2}} 
\def\D28{D_{28}^{-2}} \def\nuo{\nu_{14.6}}
\def\Tdone{T_{d,-1}} \def\Tdtwo{T_{d,-2}} \def\Tdthree{T_{d,-3}}

\date{Accepted . Received ; }

\pagerange{\pageref{firstpage}--\pageref{lastpage}} \pubyear{2012}

\label{firstpage}

\maketitle

\begin{abstract}
We investigate wheter there is any correlation between the X--ray afterglow luminosity and the prompt emission properties of a carefully selected sub-sample of bright {\it Swift} long Gamma--Ray Bursts (GRBs) nearly complete in redshift ($\sim 90\%$). Being free of selection effects (except flux limit), this sample provides the possibility to compare the rest frame physical properties of GRB prompt and afterglow emission in an unbiased way. The afterglow X--ray luminosities are computed at four different rest frame times (5 min, 1 hr, 11 hr and 24 hr after trigger) and compared with the prompt emission isotropic energy $E_{iso}$, the isotropic peak luminosity $L_{iso}$ and the rest frame peak energy $E_{peak}$. We find that the rest frame afterglow X--ray luminosity do correlate with these prompt emission quantities, but the significance of each correlation decreases over time. This result is in agreement with the idea that the GRB X-ray light curve can be described as the result of a combination of different components whose relative contribution and weight change with time, with the prompt and afterglow emission dominating at early and late time, respectively. In particular, we found evidence that the plateau and the shallow decay phase often observed in GRB X--ray light curves are powered by activity from the central engine. The existence of the $L_X-E_{iso}$ correlation at late times ($t_{rf} \geq 11$ hr) suggests a similar radiative efficiency among different bursts with on average about 6\% of the total kinetic energy powering the prompt emission.

\end{abstract}

\begin{keywords}
gamma-rays: bursts -- X--rays: general.
\end{keywords}

\section{Introduction}

Gamma-ray bursts (GRBs) are rapid, intense flashes of gamma-ray radiation occuring
at an average rate of one event per day at cosmological distances. The high energy prompt emission is followed by a broadband (X-rays to radio ranges) afterglow (Costa et al. 1997; van Paradijs et al. 1997; Frail et al. 1997; Bremer et al. 1998; Heng et al. 2008) that can be observed up to weeks and months after the onset of the event. The {\it Swift} satellite (Gehrels et al. 2004) is operative since 2004 and provided, so far, uniform observations of prompt and afterglow emission for hundreds of GRBs. Among the most interesting uses of the {\it Swift} legacy, there are the statistical studies aimed at searching for the existence of correlations among the physical parameters of GRBs. From the study of the GRB prompt properties, robust correlations among the spectral parameters of the prompt emission and its energetic and luminosity have been found (Amati et al. 2002; Yonetoku et al. 2004; Ghirlanda, Ghisellini \& Lazzati 2004). Furthermore, many studies searching for correlations among GRB prompt and afterglow emission have been presented so far in the literature, based on the comparison of the observed properties (see, e.g. Gehrels et al. 2008 and references therein) or of the rest-frame properties, selecting GRBs for which a redshift could be measured (Berger et al. 2003; Racusin et al. 2011; Kann et al. 2010; Kann et al. 2011; Nysewander et al. 2009; Ghirlanda et al. 2009; Margutti et al. 2012; Bernardini et al. 2012a,b).
Although this second approach enable to physically characterize these objects, it can be affected by observational biases, given that almost 2/3 of {\it Swift} GRBs are lacking a redshift measurement.
Such correlation studies are of key importance for the understanding of the physics of GRB emission mechanisms and of their relation to their progenitors and to the surrounding environment. 

In this paper, we investigate the existence of correlations among afterglow emission and prompt spectral properties of a carefully selected sub-sample of {\it Swift} long GRBs presented in Salvaterra et al. (2012). This sample is nearly complete in redshift ($\sim 90\%$) and, consisting of 58 GRBs, is large enough to allow significant statistical studies.

The paper is organized as follows. In section 2 we describe the properties of the GRBs composing the complete sample of Salvaterra et al. (2012). In section 3 we compare these properties and discuss our findings. Our conclusion are presented in section 4. Throughout the paper we assume a standard cosmology with $h={\Omega}_{\Lambda} = 0.7$ and ${\Omega}_{m} = 0.3$.

\section{The sample: selection and correlations}

Jakobsson et al. (2006) proposed to select among long GRBs (those with duration above $\sim 2$ s; Kouveliotou et al. 1993) detected by {\it Swift} only those with favorable observing conditions for ground-based optical follow-up\footnote{http://www.raunvis.hi.is/pja/GRBsample.html} aimed at redshift determination. This sample has a completeness in redshift of $\sim 53\%$. Salvaterra et al. (2012) restricted this sample to those events with a peak photon flux $P \geq 2.6$ ph s$^{-1}$ cm$^{-2}$, measured in the 15--150 keV energy band by the {\it Swift-}BAT. This further criterium selects 58 GRBs, 52 with a measured redshift (a completeness level of 90\%, which increases up to 95\% considerings events with some constraint on $z$). Being free of selection effects (except flux limit), such sample provides the possibility to compare the rest-frame physical properties of GRB prompt and afterglow emission in an unbiased way. 

Using the automated data products provided by the {\it Swift}/XRT light curve\footnote{http://www.swift.ac.uk/xrt\_curves/} and spectra\footnote{http://www.swift.ac.uk/xrt\_spectra/} repositories (Evans et al. 2009) we estimated the afterglow X--ray integral fluxes in the 2-10 keV rest frame common energy band and computed the corresponding rest frame X--ray luminosities at different rest frame times for all the GRBs of our sample with a measured redshift. The 2-10 keV rest frame fluxes were computed from the observed integral 0.3-10 keV unabsorbed fluxes and the measured spectral index, $\Gamma$, (that we retrieved from the in the {\it Swift}/XRT data repositories above) in the following way (see also Gehrels et al. 2008):  

\begin{equation} 
f_{X,rf}(2-10 \, {\rm{keV}}) = f_X(0.3-10 \, \rm{keV})\frac{\left({\frac{10}{1+z}}\right)^{2-\Gamma}-\left({\frac{2}{1+z}}\right)^{2-\Gamma}}{{10}^{2-\Gamma}-{0.3}^{2-\Gamma}}
\label{kcorr_eq}
\end{equation}

\noindent The X--ray light curves were fitted with power laws, broken power laws or multiply broken power laws (after removing the time intervals showing significant flaring) and then the fits where interpolated or extrapolated to the given rest frame times. A peculiar case is given by GRB\,060614, whose light curve was fitted by an exponential function plus a broken power-law (see also Mangano et al. 2007).

\subsection{Correlation Analysis}

The obtained afterglow X--ray luminosities were compared with the prompt emission isotropic energy $E_{iso}$, the isotropic peak luminosity $L_{iso}$ and the rest frame peak energy $E_{peak}$ reported in a companion paper by Nava et al. (2012) for the bursts of our sample. As a result, we obtained 46 GRBs for which all the quantities $L_X$, $E_{iso}$, $L_{iso}$ and $E_{peak}$ were determined (79\% of the sample). GRBs with limits on the redshift were not included in our analysis.

\begin{figure}
   \centering
   \includegraphics[width=84mm]{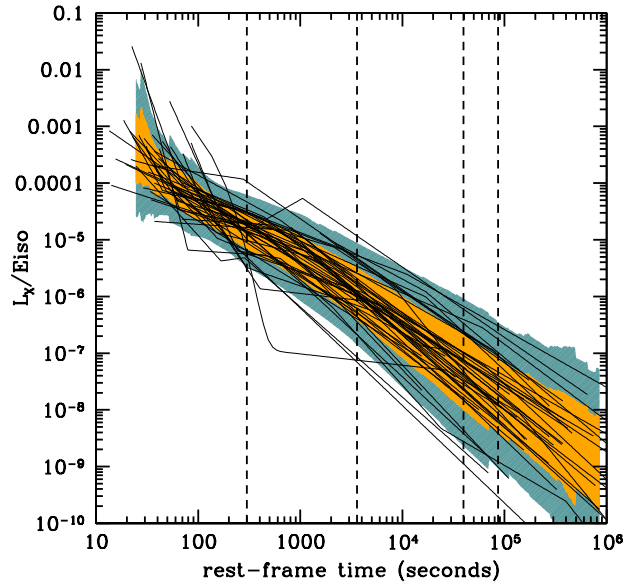}
  \caption{Best fit of the X--ray luminosity light curves of the 46 GRBs analyzed in this paper normalized to their $E_{iso}$. The X-ray luminosities were computed for each GRB in the common rest frame $2-10$ keV energy band following the precedure described in Sec. 2. The rest frame times at which we computed $L_X-E_{iso}$, $L_X-E_{peak}$ and $L_X-L_{iso}$ correlations are marked with vertical dashed lines. The light (dark) shaded area represents the $1\sigma$ ($2\sigma$) scatter around the mean value of the $L_{iso}/E_{iso}$ distribution at a given rest-frame time $t_{rf}$.
}
       \label{fig:$E_{iso}$_Lx}
\end{figure}

In Fig.~1 we show the X-ray light curves of the 46 GRBs analyzed in this paper normalized to their $E_{iso}$. This plot shows that the $E_{iso}-$normalized X-ray light curves are rather clustered, with an intrinsic scatter that changes with the rest frame time. With the aim of investigating such evolution in time between the prompt and X-ray afterglow emission, different correlations ($L_X-E_{iso}$, $L_X-L_{iso}$ and $L_X-E_{peak}$) were tested at four rest frame time. The early X-ray afterglow flux was measured at $t_{rf}= 5$ min and at $t_{rf}= 1$ hr, i.e. at the end of the prompt phase and during the expected plateau phase of the ``canonical'' X--ray light curve (Nousek et al. 2006; Zhang et al. 2006; O'Brien et al. 2006; see also Sec. 3.1), while the late time afterglow flux was measured at $t_{rf}= 11$ hr and $t_{rf}= 24$ hr (Fig.~1)\footnote{The measure of the X-ray afterglow flux at 11 and 24 hours post burst in the observer's frame is a choice commonly made in the literature for statistical studies focussed on the study of late-time ``pure'' afterglow emission (see, e.g., De Pasquale et al. 2003, Jakobsson et al. 2004; Gehrels et al. 2008; Nysewander et al. 2009; Gendre et al. 2008). For this work we kept the same time bins converting them in the rest frame.}. We computed for each case the Spearman rank correlation coefficient $r$ (Spearman 1904; Press et al. 1986) and, in order to determine the significance of the correlation, the associated null-hypothesis probability $P_{null}$. We note that, for the events of our sample, luminosities and energies can be correlated with redshift and this could give rise to spurious correlations. Indeed, for any flux-limited sample (as our sample is) there will be an inevitable and tight correlation between luminosity and redshift.
This arises because, for a fixed flux limit, only powerful sources can be detected out to great distances (see, e.g., Blundell, Rawlings \& Willot 1999 and references therein). In order to properly handle this problem the correlations between luminosities or between luminosity and energy should be examined excluding the dependence on redshift. This can be done with a partial correlation analysis. If $r_{ij}$ is the correlation coefficient between $x_i$ and $x_j$, in the case of three variables the correlation coefficient between two of them, excluding the effect of the third one is

\begin{equation} 
r_{12,3} = \frac{r_{12}-r_{13}r_{23}}{\sqrt{1-r^{2}_{13}}\sqrt{1-r^{2}_{23}}}
\label{pad_eq}
\end{equation}

\noindent (Kendall \& Stuart 1979; see also Padovani 1992) where, for our study, the coefficients 1 and 2 are $L_X$ and $E_{iso}$ (or $L_{iso}$ or $E_{peak}$), respectively, and the coefficient 3 is the redshift. We also performed a linear fit to each data distribution in logarithmic space. Given the large scatter of the data points (larger than the uncertainties on each value) and that, in principle, either $L_X$ or $E_{iso}$ (or $E_{peak}$ or $L_{iso}$) can be assumed as the independent variable, we fit the data using the ordinary least squares bisector method (Isobe et al. 1990). 

The list of GRBs included in our sample, together with their rest frame X--ray afterglow luminosities is presented in Table~1.

\section{Results and discussion}

The results of our correlation analysis and linear fitting are reported in Table 2 and plotted in Fig.~2. 
As a general trend, we note that the afterglow X--ray luminosity at early times ($t_{rf}=5$ min and $t_{rf}=1$ hr) strongly correlate with the prompt emission quantities $E_{iso}$ and $L_{iso}$ with null probabilities of less than $10^{-4}$ and small dispersion ($\sim 0.3$ dex). At later times 
($t_{rf}=11$ and $t_{rf}=24$ hr) the correlations become less strong, although still significant ($P_{null} \sim 10^{-2}$), with an increase in the dispersion (0.4--0.5 dex). The correlation between the X-ray luminosity and $E_{peak}$ seems to follow the same trend, even if in this case the significance is lower at all sampled times (although with probabilities $>$ 95\%). 
We also tested if these correlations are affected by evolutionary effects and if their slope change with redshift. Being complete in redshift, our sample represents an ideal test bench to perform this kind of check. To this end, we divided the sample into two redshift bins ($z < 1.8$ and $z > 1.8$), each one consisting of 23 events and repeated the analysis described in Sect. 2.1 (the results are reported in Table~3). No significant change (within $2\sigma$ c.l.) is found in the slopes of each individual bin with respect to the ones obtained for the whole sample, excluding any dependence on the redshift for the correlations.

\begin{figure*}
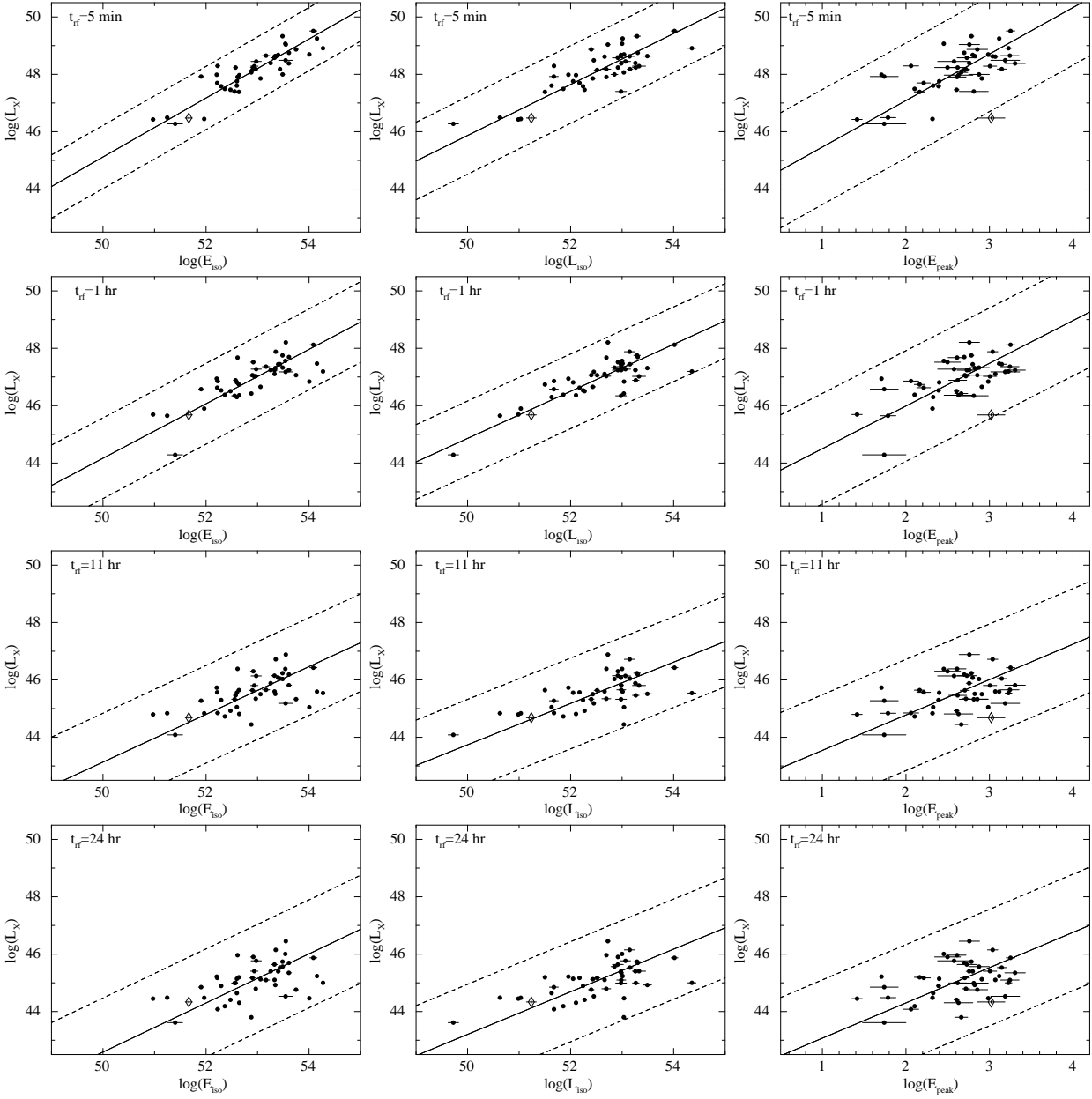

    \centering
    \includegraphics[height=5.4cm,angle=-90]{logEiso_logLx_5min_v5.ps}
    \includegraphics[height=5.4cm,angle=-90]{logLp_logLx_5min_v5.ps}
    \includegraphics[height=5.4cm,angle=-90]{logEp_logLx_5min_v5.ps}
    \includegraphics[height=5.4cm,angle=-90]{logEiso_logLx_1hr_v5.ps}
    \includegraphics[height=5.4cm,angle=-90]{logLp_logLx_1hr_v5.ps}
    \includegraphics[height=5.4cm,angle=-90]{logEp_logLx_1hr_v5.ps}
    \includegraphics[height=5.4cm,angle=-90]{logEiso_logLx_11hr_v5.ps}
    \includegraphics[height=5.4cm,angle=-90]{logLp_logLx_11hr_v5.ps}
    \includegraphics[height=5.4cm,angle=-90]{logEp_logLx_11hr_v5.ps}
    \includegraphics[height=5.4cm,angle=-90]{logEiso_logLx_24hr_v5.ps}
    \includegraphics[height=5.4cm,angle=-90]{logLp_logLx_24hr_v5.ps}
    \includegraphics[height=5.4cm,angle=-90]{logEp_logLx_24hr_v5.ps}
    \caption{The $L_X-E_{iso}$, $L_X-L_{iso}$ and $L_X-E_{peak}$ correlations studied in the present paper at different rest-frame times. The solid lines represent the best fit obtained with the method described in Sect. 2.1. The 3$\sigma$ scatter of the distribution data points lies within the dashed lines. GRB\,061021, found to be an outlier of the $E_{peak}-E_{iso}$ correlation (Nava et al. 2012), is marked with a diamond.}
       \label{fig:correlations}
\end{figure*}

\subsection{Early time correlations}

The so-called ``canonical'' X-ray light curve of GRBs (Nousek et al. 2006; Zhang et al. 2006; O'Brien et al. 2006) shows a double broken power--law profile with an initial steep decay (typically at $(t-t_0)_{obs} < 300-500$ s), followed by a plateau phase and ends (at $(t-t_0)_{obs} > 10^3-10^4$ s) with a ``normal'' afterglow decay. 
While there is general consensus on the association of the initial steep decay with the tail of the prompt emission (Kumar \& Panaitescu 2000; Tagliaferri et al. 2005) and of the normal decay with pure external forward shock afterglow emission (Sari et al. 1998; Chevalier \& Li 2000), the nature of the plateau decay phase is still debated. The usual explanation of this phase is that the observed emission is a combination of external forward shock (afterglow) and energy injection coming from late-time activity of the central engine (see, e.g., Zhang et al. 2006 and references therein). However, not all the GRBs display the canonical X-ray light curve\footnote{This is true also for the events of our complete sample, where only 18 of the 46 light curves studied in this work show the canonical  morphology.} but almost all the events can be described with a combination of some of the three different kind of power-law decays described above (see, e.g., Evans et al. 2009; Bernardini et al. 2012a). It has been shown that the steep decay phases can be considered equivalent independently of the morphology of the light curve where they are observed and that this is also valid for the normal decay phases (Evans et al. 2009; Bernardini et al. 2012a). For the sake of simplicity, in the following we will consider for our purposes also the plateau phase of the canonical X--ray light curves and the shallow decay frequently observed in the broken power-law X--ray light curves as equivalent (as also suggested by Bernardini et al. 2012a). 

As discussed above, we found the strongest correlations when we compare the early time X-ray luminosity at $t_{rf}=5$ min with $E_{iso}$ and $L_{iso}$. We plot again these early-time correlations in Fig.~3 using different markers for the different light curve decay phases (steep, plateau/shallow, normal). At this epoch almost 2/3 of the X-ray light curves are in the plateau/shallow decay phase (29 events out of 46), two shows still a steep decay and the remaining (14/46) display the normal decay\footnote{We are not considering here GRB\,060614, which light curve shows an odd profile (see Sect. 2).}. This result clearly shows that the plateau/shallow phase, as the steep decay, is likely dominated by the central engine activity, as suggested by Zhang et al. (2006). This conclusion is further supported by the fact that at later times the fraction of events showing the normal afterglow decay increases significantly (80\% at $t_{rf}=1$ hr and more than 90\% at $t_{rf} \geq 11$ hr) together with the dispersion of the correlations. This might indicate the rise of an additional emission component, afterglow dominated. Such a trend is also clear from Fig. 1, where we plotted the X-ray light curves of all 46 GRBs normalized by their $E_{iso}$.
On the other hand, Fig.~3 tells us also that a significant fraction (30\%) of GRBs of our sample shows a normal afterglow decay already at $t_{rf}=5$ min. This provides a lower limit for the rest frame time when the onset of the afterglow occurred, and implies Lorenz factors of the order of $\Gamma\geq200$ for these events. GRBs with such large values of $\Gamma$ are expected to be highly energetic (Sari \& Piran 1999; Panaitescu \& Kumar 2000; M\'esz\'aros 2006; Ghirlanda et al. 2012). To test this we compared the distributions of $E_{iso}$ of the events showing a normal afterglow decay at $t_{rf}=5$ min with those showing a plateau (or a steep decay). Althought the value of $E_{iso}$ of GRBs showing a normal decay at early time is on average higher (by a factor of $\sim 3$), a Kolmogorv-Smirnov (K-S) test gives a 22\% probability, likely indicating that the two distributions are drawn from the same population. A similar conclusion can be drawn from the same K-S test performed on the distributions of $L_{iso}$, which gives a probability of 70\%\footnote{Also in this case, the average value of $L_{iso}$ is a factor of $\sim 3$ higher for GRBs showing a normal decay at $t_{rf}=5$ min.}. 

Combining the slopes that we found in our correlations at $t_{rf}=5$ min we are in good agreement with the corresponding slopes of the $E_{peak}-E_{iso}$ and $E_{peak}-L_{iso}$ correlations (Amati et al. 2002; Yonetoku et al. 2004). This is in line with the work presented by Nava et al. (2012) where it is shown that these two correlations hold also for our complete sample of GRBs. In their analysis, Nava et al. (2012) pointed out the presence of one outlier (GRB\,061021) of the $E_{peak}-E_{iso}$ correlation. Indeed, as can be seen in Fig.~2, we find that GRB\,061021 is consistent with the $L_X-E_{iso}$ and $L_X-L_{iso}$ correlations at any time, while it falls outside the $3\sigma$ region of the $L_X-E_{peak}$ correlation. We therefore argue that the inconsistency of GRB\,061021 with the $E_{peak}-E_{iso}$ correlation is likely due to the hardness of its prompt emission spectrum (high $E_{peak}$). 
To this regard, we note that, being our sample limited to bright events (Sec. 2) faint GRBs like GRB\,980425 and GRB\,031203 are not included. These two peculiar GRBs are well known outliers of both the $E_{peak}-E_{iso}$ and $E_{peak}-L_{iso}$ correlation (while they are consistent with the three parameter $E_{\gamma,iso}-E_{peak}-E_{X,iso}$ correlation recently reported by Bernardini et al. 2012b and Margutti et al. 2012).

\begin{figure}
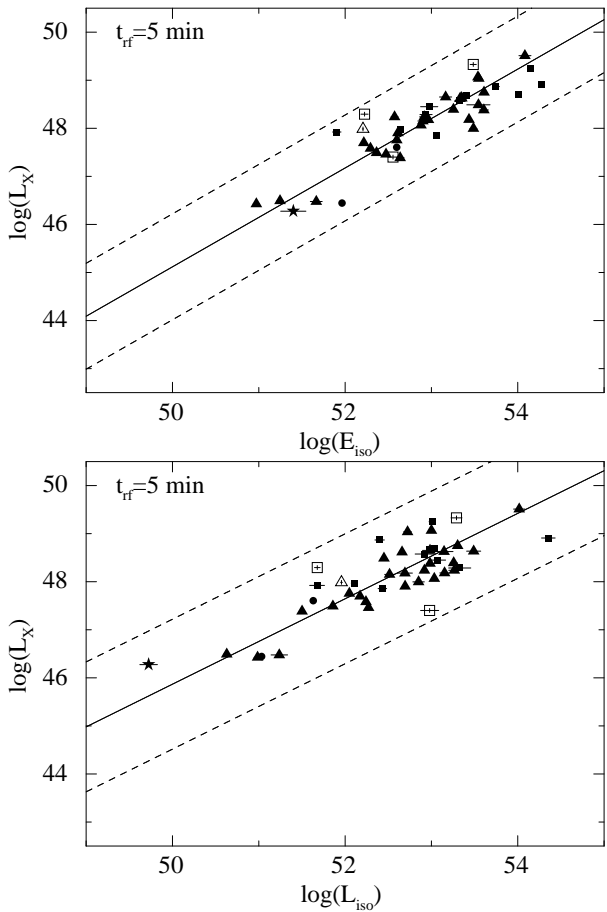

   \centering
    \includegraphics[width=6.cm,angle=-90]{logEiso_logLx_5min_classes2_v4.ps}
    \includegraphics[width=6.cm,angle=-90]{logLp_logLx_5min_classes2_v4.ps}
  \caption{Same as Fig. 2. The different X--ray light curve decay phases are indicated: steep decay (dots), plateau/shallow decay (triangles), normal decay (squares) and odd (GRB\,060614; star). Empty symbols indicate extrapolated luminosities.}
       \label{fig:classes}
\end{figure}

\subsection{Late time correlations}

It can be shown (see, e.g., Painatescu \& Kumar 2002; Berger et al. 2003) that for typical afterglow parameters the X-ray band lies above the synchrotron cooling frequency $\nu_c$ at relatively late times (few hours after the burst). Furthermore, at frequencies above $\nu_c$ the value of the afterglow flux is independent of the circumburst density medium and, for burst with a known luminosity distance, is proportional to $\epsilon_e \epsilon_B E_{K,iso}$, where $\epsilon_e$ and $\epsilon_B$ are the shock energy carried by the electrons and by the magnetic field, respectively, and $E_{K,iso}$ is the total isotropic kinetic energy of the fireball (Sari et al. 1998; Kumar 2000; Panaitescu \& Kumar 2000). Therefore, the late time X--ray luminosity can be considered as a robust proxy of $E_{K,iso}$ (Freedman \& Waxman 2001; Berger et al. 2003; Berger 2007; Nysewander et al. 2009). In light of this, the correlations between the afterglow X--ray luminosity in the 2--10 keV rest frame energy band and $E_{iso}$ we found at $t_{rf} \geq 11$ hr (Table 2) can be used to constrain the radiative efficiency of the GRB prompt emission. 
From Eq. B9 of Panaitescu \& Kumar (2000), for GRBs with known luminosity distance, the X--ray luminosity in the 2--10 keV rest frame energy band at $T_{rf}=1$ d is:

\begin{equation}
L(2-10 \, {\rm{keV}}) \propto \epse1^{p-1} \eBtwo^{\frac{p-2}{4}} E_{53}^{\frac{p+2}{4}};\    
\label{lum_rf}
\end{equation}

\noindent where $E_{K,iso}~=~10^{53}E_{53}$ erg, $\epsilon_e~=~10^{-1}\epse1$, $\epsilon_B~=~10^{-2}\eBtwo$ and $p$ is the electron energy distribution index (for GRB afterglows we have $p\sim 2.1-2.5$; Chevalier \& Li 1999). Replacing the rest-frame X--ray luminosities at $t_{rf} = 24$ hr reported in Table~1 in Eq.~\ref{lum_rf} we can thus compute the total kinetic energy for each GRB of our sample and estimate the ratio $E_{iso}/E_{K,iso}$, i.e. the radiative efficiency of the GRB prompt emission. For typical $\epsilon_e$ and $\epsilon_B$ values of $10^{-1}$ and $10^{-2}$, respectively, and assuming $p \sim 2.3$ we find an average $E_{iso}/E_{K,iso}=0.06$ ($\sigma=0.14$), in line with the values presented in past works for long and short GRBs (Freedman \& Waxman 2001; Berger et al. 2003; Granot et al. 2006; Zhang et al. 2007; Berger 2007; Nakar 2007; Nysewander et al. 2009). The distribution of the $E_{iso}/E_{K,iso}$ values for the GRBs of our sample is shown in Fig.~4.

\begin{figure}
   \centering
    \includegraphics[width=6.cm,angle=-90]{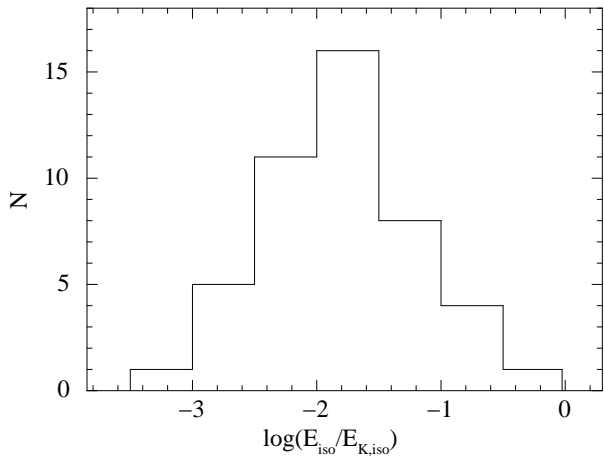}
  \caption{A histogram of the ratio between the prompt emission isotropic energy ($E_{iso}$) and the total kinetic energy ($E_{K,iso}$) of the GRBs of our sample.}
       \label{fig:eta_dist}
\end{figure}

\subsection{The contribution of prompt and afterglow emission to the GRB X--ray light curve}

According to the standard GRB fireball model (M\'esz\'aros 2002 and references therein) a central engine is supposed to produce a number of relativistic internal shocks, which interact with each other giving rise to the prompt high--energy emission, while a broadband afterglow emission is expected to arise from an external shock when the relativistically expanding fireball is decelerated by the surrounding medium. 
The observed decrease of the significance of each correlation with time (Table 2) finds a natural explanation if we assume that the GRB X-ray light curve is the result of a combination of different components whose relative contribution and weight change with time (see, e.g., Willingale et al. 2007), with the prompt (internal shock) and afterglow (external shock) emission dominating at early and late time, respectively. As a result, the early X-ray luminosity (being dominated by the prompt emission) correlate well with the prompt emission quantites $E_{iso}$, $L_{iso}$ and $E_{peak}$. On the other hand, at late time, the increase in the scatter of the $L_X-E_{iso}$, $L_X-L_{iso}$ and $L_X-E_{peak}$ correlations can be interpreted as due to the rising of an additional component (the afterglow).

\section{Conclusions}

The statistical study of the rest-frame properties of GRBs gives the best opportunity to characterize the physics of these events, although such studies are often biased by the fact that almost 2/3 of GRBs are lacking a redshift measurement. In this paper, working with a sample of GRBs complete in redshift ($\sim 90\%$), we investigated the existence of correlations among the GRB X--ray afterglow luminosities and rest-frame prompt emission properties in an unbiased way. We tested the correlations between luminosities or between luminosity and energy performing a partial correlation analysis against the common dependence on the redshift (see Sect. 2.1) and obtained the following main results:

- the afterglow X--ray luminosity $L_X$ at early times ($t_{rf}=5$ min and $t_{rf}=1$ hr) strongly correlate with the prompt emission isotropic energy ($E_{iso}$) and the peak luminosity ($L_{iso}$). At later times ($t_{rf}=11$ hr and $t_{rf}=24$ hr) the $L_X-E_{iso}$ and $L_X-L_{iso}$ correlations become weaker (but still significant). A similar trend is observed between $L_X$ and $E_{peak}$, even if in this case the significance of the correlation is lower at all times (although with correlation probabilities $>$ 95\%). 

- the strongest correlations are found comparing the early time X-ray luminosity at $t_{rf}=5$ min with $E_{iso}$ and $L_{iso}$. At this epoch the majority of the X-ray light curves display a steep or a plateau/shallow decay. This suggests that the initial steep decay and the plateau (shallow decay) phase observed in GRB X--ray light curves are powered by activity from the central engine;

- GRBs showing the normal afterglow decay already at $t_{rf}=5$ min, althought expected to have high value of the Lorenz factor, do not show a significant excess in their $E_{iso}$ and $L_{iso}$ with respect to the other events;

- the existence of the $L_X-E_{iso}$ correlation at late times suggests a similar radiative efficiency ($\sim$6\%) among different bursts, where brighter bursts have on average more kinetic energy;

- the resulting slopes of our correlation are in agreement with the Amati and Yonetoku correlations (see also Nava et al. 2012). The inconsistency of GRB\,061021 with the $E_{peak}-E_{iso}$ correlation is likely due to the hardness of its prompt emission spectrum (high $E_{peak}$);

- the decrease of significance of these correlations with time indicates that the early X--ray luminosity is still dominated by the prompt emission, while at late times the most significant contribution to the X--ray luminosity is given by the external shock afterglow emission.

\begin{table*}
\caption{X--ray luminosities in the 2--10 keV rest frame energy band computed at four different rest frame times (5 min, 1 hr, 11 hr and 24 hr) for the GRBs studied in this paper (see Sect. 2 for details). Errors are at the 90\% confidence level.
}
\centering
\begin{tabular}{cccccccccc} \hline 
GRB     &  $z$      &  $L_{X,5}$	      &   $L_{X,5}$ err 	&   $L_{X,1}$		  &   $L_{X,1}$ err	    & $L_{X,11}$	      &   $L_{X,11}$ err 	& $L_{X,24}$		  &   $L_{X,24}$ err	       \\
        &           &  erg s$^{-1}$	      &  erg s$^{-1}$		&  erg s$^{-1}$ 	  &  erg s$^{-1}$	    &  erg s$^{-1}$	      &  erg s$^{-1}$		&  erg s$^{-1}$ 	  &  erg s$^{-1}$	       \\ \hline
050318  & 1.440     &  $1.966 \times 10^{48}$ &  $3.932 \times 10^{47}$ &  $7.197 \times 10^{46}$ &  $1.583 \times 10^{46}$ &  $7.054 \times 10^{44}$ &  $1.622 \times 10^{44}$ &  $1.202 \times 10^{44}$ &  $3.245 \times 10^{43}$    \\
050401  & 2.900     &  $5.624 \times 10^{48}$ &  $1.125 \times 10^{48}$ &  $4.971 \times 10^{47}$ &  $1.094 \times 10^{47}$ &  $1.543 \times 10^{46}$ &  $3.549 \times 10^{45}$ &  $4.959 \times 10^{45}$ &  $1.339 \times 10^{45}$    \\
050416A & 0.650     &  $2.676 \times 10^{46}$ &  $5.352 \times 10^{45}$ &  $4.933 \times 10^{45}$ &  $1.085 \times 10^{45}$ &  $6.252 \times 10^{44}$ &  $1.438 \times 10^{44}$ &  $2.813 \times 10^{44}$ &  $7.595 \times 10^{43}$    \\
050525A & 0.610     &  $3.113 \times 10^{47}$ &  $6.226 \times 10^{46}$ &  $2.399 \times 10^{46}$ &  $5.278 \times 10^{45}$ &  $5.326 \times 10^{44}$ &  $1.225 \times 10^{44}$ &  $1.534 \times 10^{44}$ &  $4.142 \times 10^{43}$    \\
050922C & 2.200     &  $1.725 \times 10^{48}$ &  $3.450 \times 10^{47}$ &  $7.660 \times 10^{46}$ &  $1.685 \times 10^{46}$ &  $2.880 \times 10^{45}$ &  $6.624 \times 10^{44}$ &  $9.855 \times 10^{44}$ &  $2.661 \times 10^{44}$    \\
060206  & 4.050     &  $8.070 \times 10^{47}$ &  $1.614 \times 10^{47}$ &  $4.733 \times 10^{47}$ &  $1.041 \times 10^{47}$ &  $2.430 \times 10^{46}$ &  $5.589 \times 10^{45}$ &  $9.204 \times 10^{45}$ &  $2.485 \times 10^{45}$    \\
060210  & 3.910     &  $1.092 \times 10^{49}$ &  $2.184 \times 10^{48}$ &  $1.602 \times 10^{48}$ &  $3.524 \times 10^{47}$ &  $7.653 \times 10^{46}$ &  $1.760 \times 10^{46}$ &  $2.832 \times 10^{46}$ &  $7.646 \times 10^{45}$    \\ 
060306  & 3.500     &  $1.731 \times 10^{48}$ &  $3.462 \times 10^{47}$ &  $3.284 \times 10^{47}$ &  $7.225 \times 10^{46}$ &  $1.993 \times 10^{46}$ &  $4.584 \times 10^{45}$ &  $7.976 \times 10^{45}$ &  $2.154 \times 10^{45}$    \\
060614  & 0.130     &  $1.883 \times 10^{46}$ &  $3.766 \times 10^{45}$ &  $1.928 \times 10^{44}$ &  $4.242 \times 10^{43}$ &  $1.214 \times 10^{44}$ &  $2.792 \times 10^{43}$ &  $4.114 \times 10^{43}$ &  $1.111 \times 10^{43}$    \\ 
060814  & 1.920     &  $9.902 \times 10^{47}$ &  $1.980 \times 10^{47}$ &  $2.119 \times 10^{47}$ &  $4.662 \times 10^{46}$ &  $1.073 \times 10^{46}$ &  $2.468 \times 10^{45}$ &  $3.671 \times 10^{45}$ &  $9.912 \times 10^{44}$    \\
060908  & 1.880     &  $9.362 \times 10^{47}$ &  $1.872 \times 10^{47}$ &  $2.302 \times 10^{46}$ &  $5.064 \times 10^{45}$ &  $6.495 \times 10^{44}$ &  $1.494 \times 10^{44}$ &  $2.023 \times 10^{44}$ &  $5.462 \times 10^{43}$    \\
060927  & 5.470     &  $1.166 \times 10^{48}$ &  $2.332 \times 10^{47}$ &  $2.644 \times 10^{46}$ &  $5.817 \times 10^{45}$ &  $2.794 \times 10^{44}$ &  $6.426 \times 10^{43}$ &  $6.314 \times 10^{43}$ &  $1.705 \times 10^{43}$    \\
061007  & 1.260     &  $4.946 \times 10^{48}$ &  $9.892 \times 10^{47}$ &  $6.865 \times 10^{46}$ &  $1.510 \times 10^{46}$ &  $1.116 \times 10^{45}$ &  $2.567 \times 10^{44}$ &  $2.904 \times 10^{44}$ &  $7.841 \times 10^{43}$    \\
061021  & 0.350     &  $2.999 \times 10^{46}$ &  $5.998 \times 10^{45}$ &  $4.761 \times 10^{45}$ &  $1.047 \times 10^{45}$ &  $4.823 \times 10^{44}$ &  $1.109 \times 10^{44}$ &  $2.153 \times 10^{44}$ &  $5.813 \times 10^{43}$    \\ 
061121  & 1.310     &  $1.524 \times 10^{48}$ &  $3.048 \times 10^{47}$ &  $2.788 \times 10^{47}$ &  $6.134 \times 10^{46}$ &  $1.108 \times 10^{46}$ &  $2.548 \times 10^{45}$ &  $3.396 \times 10^{45}$ &  $9.169 \times 10^{44}$    \\
061222A & 2.090     &  $4.237 \times 10^{48}$ &  $8.474 \times 10^{47}$ &  $7.602 \times 10^{47}$ &  $1.672 \times 10^{47}$ &  $5.253 \times 10^{46}$ &  $1.208 \times 10^{46}$ &  $1.410 \times 10^{46}$ &  $3.807 \times 10^{45}$    \\
070521  & 1.350     &  $1.512 \times 10^{48}$ &  $3.024 \times 10^{47}$ &  $1.068 \times 10^{47}$ &  $2.350 \times 10^{46}$ &  $2.207 \times 10^{45}$ &  $5.076 \times 10^{44}$ &  $6.208 \times 10^{44}$ &  $1.676 \times 10^{44}$    \\
071020  & 2.150     &  $1.929 \times 10^{48}$ &  $3.858 \times 10^{47}$ &  $1.051 \times 10^{47}$ &  $2.312 \times 10^{46}$ &  $6.381 \times 10^{45}$ &  $1.468 \times 10^{45}$ &  $2.553 \times 10^{45}$ &  $6.893 \times 10^{44}$    \\
071117  & 1.330     &  $2.524 \times 10^{47}$ &  $5.048 \times 10^{46}$ &  $2.207 \times 10^{46}$ &  $4.855 \times 10^{45}$ &  $2.111 \times 10^{45}$ &  $4.855 \times 10^{44}$ &  $9.802 \times 10^{44}$ &  $2.647 \times 10^{44}$    \\
080319B & 0.940     &  $1.774 \times 10^{49}$ &  $3.548 \times 10^{48}$ &  $2.945 \times 10^{47}$ &  $6.479 \times 10^{46}$ &  $3.903 \times 10^{45}$ &  $8.977 \times 10^{44}$ &  $1.716 \times 10^{45}$ &  $4.633 \times 10^{44}$    \\
080319C & 1.950     &  $4.485 \times 10^{48}$ &  $8.970 \times 10^{47}$ &  $2.291 \times 10^{47}$ &  $5.040 \times 10^{46}$ &  $4.512 \times 10^{45}$ &  $1.038 \times 10^{45}$ &  $1.250 \times 10^{45}$ &  $3.375 \times 10^{44}$    \\ 
080413B & 1.100     &  $4.996 \times 10^{47}$ &  $9.992 \times 10^{46}$ &  $4.230 \times 10^{46}$ &  $9.306 \times 10^{45}$ &  $3.678 \times 10^{45}$ &  $8.459 \times 10^{44}$ &  $1.488 \times 10^{45}$ &  $4.018 \times 10^{44}$    \\
080603B & 2.690     &  $2.816 \times 10^{48}$ &  $5.632 \times 10^{47}$ &  $1.873 \times 10^{47}$ &  $4.121 \times 10^{46}$ &  $1.377 \times 10^{46}$ &  $3.167 \times 10^{45}$ &  $5.865 \times 10^{45}$ &  $1.584 \times 10^{45}$    \\
080605  & 1.640     &  $4.327 \times 10^{48}$ &  $8.654 \times 10^{47}$ &  $2.052 \times 10^{47}$ &  $4.514 \times 10^{46}$ &  $3.258 \times 10^{45}$ &  $7.493 \times 10^{44}$ &  $8.409 \times 10^{44}$ &  $2.270 \times 10^{44}$    \\
080607  & 3.040     &  $8.135 \times 10^{48}$ &  $1.627 \times 10^{48}$ &  $1.560 \times 10^{47}$ &  $3.432 \times 10^{46}$ &  $3.464 \times 10^{45}$ &  $7.967 \times 10^{44}$ &  $9.975 \times 10^{44}$ &  $2.693 \times 10^{44}$    \\
080721  & 2.590     &  $3.240 \times 10^{49}$ &  $6.480 \times 10^{48}$ &  $1.320 \times 10^{48}$ &  $2.904 \times 10^{47}$ &  $2.664 \times 10^{46}$ &  $6.127 \times 10^{45}$ &  $7.436 \times 10^{45}$ &  $2.008 \times 10^{45}$    \\
080804  & 2.200     &  $7.173 \times 10^{47}$ &  $1.435 \times 10^{47}$ &  $4.539 \times 10^{46}$ &  $9.986 \times 10^{45}$ &  $3.181 \times 10^{45}$ &  $7.316 \times 10^{44}$ &  $1.334 \times 10^{45}$ &  $3.602 \times 10^{44}$    \\
080916A & 0.690     &  $2.784 \times 10^{46}$ &  $5.568 \times 10^{45}$ &  $7.932 \times 10^{45}$ &  $1.745 \times 10^{45}$ &  $6.859 \times 10^{44}$ &  $1.578 \times 10^{44}$ &  $3.015 \times 10^{44}$ &  $8.141 \times 10^{43}$    \\ 
081007  & 0.530     &  $3.102 \times 10^{46}$ &  $6.204 \times 10^{45}$ &  $4.460 \times 10^{45}$ &  $9.812 \times 10^{44}$ &  $6.888 \times 10^{44}$ &  $1.584 \times 10^{44}$ &  $3.050 \times 10^{44}$ &  $8.235 \times 10^{43}$    \\
081121  & 2.510     &  $2.135 \times 10^{49}$ &  $4.270 \times 10^{48}$ &  $5.657 \times 10^{47}$ &  $1.245 \times 10^{47}$ &  $1.715 \times 10^{46}$ &  $3.945 \times 10^{45}$ &  $5.468 \times 10^{45}$ &  $1.476 \times 10^{45}$    \\
081203A & 2.100     &  $3.081 \times 10^{48}$ &  $6.162 \times 10^{47}$ &  $1.508 \times 10^{47}$ &  $3.318 \times 10^{46}$ &  $1.519 \times 10^{45}$ &  $3.494 \times 10^{44}$ &  $3.379 \times 10^{44}$ &  $9.123 \times 10^{43}$    \\
081221  & 2.260     &  $1.167 \times 10^{49}$ &  $2.334 \times 10^{48}$ &  $3.630 \times 10^{47}$ &  $7.986 \times 10^{46}$ &  $2.416 \times 10^{46}$ &  $5.557 \times 10^{45}$ &  $1.013 \times 10^{46}$ &  $2.735 \times 10^{45}$    \\
081222  & 2.770     &  $4.728 \times 10^{48}$ &  $9.456 \times 10^{47}$ &  $2.776 \times 10^{47}$ &  $6.107 \times 10^{46}$ &  $1.186 \times 10^{46}$ &  $2.728 \times 10^{45}$ &  $2.498 \times 10^{45}$ &  $6.745 \times 10^{44}$    \\
090102  & 1.550     &  $4.146 \times 10^{48}$ &  $8.292 \times 10^{47}$ &  $1.270 \times 10^{47}$ &  $2.794 \times 10^{46}$ &  $3.943 \times 10^{45}$ &  $9.069 \times 10^{44}$ &  $1.267 \times 10^{45}$ &  $3.421 \times 10^{44}$    \\
090424  & 0.540     &  $5.712 \times 10^{47}$ &  $1.142 \times 10^{47}$ &  $6.465 \times 10^{46}$ &  $1.422 \times 10^{46}$ &  $3.566 \times 10^{45}$ &  $8.202 \times 10^{44}$ &  $1.383 \times 10^{45}$ &  $3.734 \times 10^{44}$    \\ 
090715B & 3.000     &  $3.764 \times 10^{48}$ &  $7.528 \times 10^{47}$ &  $1.724 \times 10^{47}$ &  $3.793 \times 10^{46}$ &  $1.403 \times 10^{46}$ &  $3.227 \times 10^{45}$ &  $4.369 \times 10^{45}$ &  $1.180 \times 10^{45}$    \\
090812  & 2.450     &  $2.409 \times 10^{48}$ &  $4.818 \times 10^{47}$ &  $1.735 \times 10^{47}$ &  $3.817 \times 10^{46}$ &  $6.525 \times 10^{45}$ &  $1.501 \times 10^{45}$ &  $2.232 \times 10^{45}$ &  $6.026 \times 10^{44}$    \\
090926B & 1.240     &  $4.025 \times 10^{47}$ &  $8.050 \times 10^{46}$ &  $1.991 \times 10^{46}$ &  $4.380 \times 10^{45}$ &  $1.125 \times 10^{45}$ &  $2.588 \times 10^{44}$ &  $4.396 \times 10^{44}$ &  $1.187 \times 10^{44}$    \\
091018  & 0.970     &  $8.393 \times 10^{47}$ &  $1.679 \times 10^{47}$ &  $3.749 \times 10^{46}$ &  $8.248 \times 10^{45}$ &  $1.879 \times 10^{45}$ &  $4.322 \times 10^{44}$ &  $7.062 \times 10^{44}$ &  $1.907 \times 10^{44}$    \\
091020  & 1.710     &  $1.419 \times 10^{48}$ &  $2.838 \times 10^{47}$ &  $1.151 \times 10^{47}$ &  $2.532 \times 10^{46}$ &  $4.329 \times 10^{45}$ &  $9.957 \times 10^{44}$ &  $1.481 \times 10^{45}$ &  $3.999 \times 10^{44}$    \\
091127  & 0.490     &  $9.707 \times 10^{47}$ &  $1.941 \times 10^{47}$ &  $8.700 \times 10^{46}$ &  $1.914 \times 10^{46}$ &  $5.367 \times 10^{45}$ &  $1.234 \times 10^{45}$ &  $1.646 \times 10^{45}$ &  $4.444 \times 10^{44}$    \\
091208B & 1.060     &  $3.825 \times 10^{47}$ &  $7.650 \times 10^{46}$ &  $3.422 \times 10^{46}$ &  $7.528 \times 10^{45}$ &  $1.980 \times 10^{45}$ &  $4.554 \times 10^{44}$ &  $7.799 \times 10^{44}$ &  $2.106 \times 10^{44}$    \\ 
100621A & 0.540     &  $2.438 \times 10^{47}$ &  $4.876 \times 10^{46}$ &  $5.484 \times 10^{46}$ &  $1.206 \times 10^{46}$ &  $4.382 \times 10^{45}$ &  $1.008 \times 10^{45}$ &  $1.558 \times 10^{45}$ &  $4.207 \times 10^{44}$    \\
100728B & 2.106     &  $2.891 \times 10^{47}$ &  $5.782 \times 10^{46}$ &  $3.204 \times 10^{46}$ &  $7.049 \times 10^{45}$ &  $8.411 \times 10^{44}$ &  $1.935 \times 10^{44}$ &  $2.559 \times 10^{44}$ &  $6.909 \times 10^{43}$    \\
110205A & 2.220     &  $7.380 \times 10^{48}$ &  $1.476 \times 10^{48}$ &  $1.160 \times 10^{47}$ &  $2.552 \times 10^{46}$ &  $2.126 \times 10^{45}$ &  $4.890 \times 10^{44}$ &  $5.752 \times 10^{44}$ &  $1.553 \times 10^{44}$    \\
110503A & 1.613     &  $2.480 \times 10^{48}$ &  $4.960 \times 10^{47}$ &  $1.733 \times 10^{47}$ &  $3.813 \times 10^{46}$ &  $7.646 \times 10^{45}$ &  $1.759 \times 10^{45}$ &  $2.516 \times 10^{45}$ &  $6.793 \times 10^{44}$    \\ \hline
\hline
\end{tabular}   
\label{tab_log1}
\end{table*}

\begin{table*}
   \centering
\caption{Correlation fits and coefficients. Data distributions were fitted with the function $y = 10^Ax^B$ (see Sect. 2.1 for details). Errors are at $1\sigma$ confidence level.}
\begin{tabular}{cccccc}
\hline
Correlation	 & A		     & B	       & $r$  & $P_{null}$	       & Dispersion \\
                 & 		     & 		       &      & 		       &	    \\ \hline
$L_{X,5}$ $vs.$ $E_{iso}$    &  $-6.31 \pm 2.70$ & $1.03 \pm 0.04$ & 0.80 & $1.44 \times 10^{-12}$ & 0.256   \\
$L_{X,1}$ $vs.$ $E_{iso}$    &  $-3.30 \pm 3.69$ & $0.95 \pm 0.05$ & 0.63 & $1.65 \times 10^{-6}$  & 0.341   \\
$L_{X,11}$ $vs.$ $E_{iso}$   &  $ 1.51 \pm 5.55$ & $0.83 \pm 0.07$ & 0.36 & $1.45 \times 10^{-2}$  & 0.436   \\
$L_{X,24}$ $vs.$ $E_{iso}$   &  $-0.21 \pm 6.92$ & $0.86 \pm 0.08$ & 0.31 & $3.37 \times 10^{-2}$  & 0.477   \\
$L_{X,5}$ $vs.$ $L_{iso}$    &  $ 1.44 \pm 3.06$ & $0.89 \pm 0.04$ & 0.59 & $1.06 \times 10^{-6}$  & 0.337   \\
$L_{X,1}$ $vs.$ $L_{iso}$    &  $ 3.85 \pm 3.04$ & $0.82 \pm 0.04$ & 0.51 & $2.47 \times 10^{-4}$  & 0.335   \\
$L_{X,11}$ $vs.$ $L_{iso}$   &  $ 7.70 \pm 4.35$ & $0.72 \pm 0.05$ & 0.36 & $1.38 \times 10^{-2}$  & 0.428   \\
$L_{X,24}$ $vs.$ $L_{iso}$   &  $ 6.06 \pm 4.98$ & $0.74 \pm 0.06$ & 0.35 & $1.93 \times 10^{-2}$  & 0.467   \\
$L_{X,5}$ $vs.$ $E_{peak}$   &  $43.84 \pm 6.74$ & $1.62 \pm 0.16$ & 0.49 & $5.98 \times 10^{-4}$  & 0.351   \\
$L_{X,1}$ $vs.$ $E_{peak}$   &  $43.01 \pm 6.74$ & $1.49 \pm 0.15$ & 0.36 & $1.40 \times 10^{-2}$  & 0.358   \\
$L_{X,11}$ $vs.$ $E_{peak}$  &  $42.31 \pm 7.46$ & $1.23 \pm 0.16$ & 0.15 & $3.31 \times 10^{-1}$  & 0.407   \\
$L_{X,24}$ $vs.$ $E_{peak}$  &  $41.84 \pm 7.53$ & $1.23 \pm 0.17$ & 0.11 & $4.71 \times 10^{-1}$  & 0.427   \\ \hline
\hline
\end{tabular}
\label{tab:log_corr}
\end{table*}

\begin{table*}
   \centering
\caption{Correlation fits and coefficients for the GRBs of our sample divided into two redshift bins ($z < 1.8$ and $z> 1.8 $; see Sec. 3 for details). Data distributions were fitted with the function $y = 10^Ax^B$ (see Sect. 2.1 for details). Errors are at $1\sigma$ confidence level.}
\begin{tabular}{cccccc}
\hline
Correlation	             & A		 & B		   & $r$  & $P_{null}$  	   & Dispersion \\
                             &  		 &		   &	  &			   &		\\ \hline
\multicolumn{6}{|c|}{$z < 1.8$}                                                                                 \\ \hline
$L_{X,5}$ $vs.$ $E_{iso}$    &  $-5.14 \pm 3.97$ & $1.01 \pm 0.05$ & 0.68 & $2.72 \times 10^{-4}$  & 0.284	\\
$L_{X,1}$ $vs.$ $E_{iso}$    &  $-0.45 \pm 4.98$ & $0.89 \pm 0.06$ & 0.73 & $5.92 \times 10^{-5}$  & 0.351	\\
$L_{X,11}$ $vs.$ $E_{iso}$   &  $13.91 \pm 6.12$ & $0.60 \pm 0.06$ & 0.51 & $1.38 \times 10^{-2}$  & 0.355	\\
$L_{X,24}$ $vs.$ $E_{iso}$   &  $12.13 \pm 7.25$ & $0.62 \pm 0.07$ & 0.49 & $1.91 \times 10^{-2}$  & 0.399	\\
$L_{X,5}$ $vs.$ $L_{iso}$    &  $ 1.84 \pm 4.26$ & $0.88 \pm 0.05$ & 0.66 & $5.25 \times 10^{-4}$  & 0.350	\\
$L_{X,1}$ $vs.$ $L_{iso}$    &  $ 5.88 \pm 3.41$ & $0.78 \pm 0.04$ & 0.70 & $1.84 \times 10^{-4}$  & 0.283	\\
$L_{X,11}$ $vs.$ $L_{iso}$   &  $18.89 \pm 4.54$ & $0.51 \pm 0.04$ & 0.53 & $9.84 \times 10^{-3}$  & 0.290	\\
$L_{X,24}$ $vs.$ $L_{iso}$   &  $17.66 \pm 5.27$ & $0.52 \pm 0.04$ & 0.58 & $4.21 \times 10^{-3}$  & 0.334	\\
$L_{X,5}$ $vs.$ $E_{peak}$   &  $44.08 \pm 10.29$ & $1.50 \pm 0.23$ & 0.37 & $9.09 \times 10^{-2}$  & 0.417	\\
$L_{X,1}$ $vs.$ $E_{peak}$   &  $43.30 \pm 9.86$ & $1.35 \pm 0.20$ & 0.37 & $8.88 \times 10^{-2}$  & 0.403	\\
$L_{X,11}$ $vs.$ $E_{peak}$  &  $43.02 \pm 11.07$ & $0.93 \pm 0.15$ & 0.22 & $3.39 \times 10^{-1}$  & 0.364	\\
$L_{X,24}$ $vs.$ $E_{peak}$  &  $42.54 \pm 11.27$ & $0.94 \pm 0.16$ & 0.30 & $1.80 \times 10^{-1}$  & 0.374	\\ 
\hline
\multicolumn{6}{|c|}{$z > 1.8$}                                                                                 \\ \hline
$L_{X,5}$ $vs.$ $E_{iso}$    &  $ -7.77 \pm 6.18$ &  $1.06 \pm 0.08$ & 0.77 & $9.76 \times 10^{-6}$ & 0.230	\\
$L_{X,1}$ $vs.$ $E_{iso}$    &  $-8.43 \pm 12.55$ &  $1.05 \pm 0.08$ & 0.44 & $3.82 \times 10^{-2}$  & 0.333	\\
$L_{X,11}$ $vs.$ $E_{iso}$   &  $-17.12 \pm 30.94$ & $1.18 \pm 0.24$ & 0.28 & $2.18 \times 10^{-1}$  & 0.438	\\
$L_{X,24}$ $vs.$ $E_{iso}$   &  $-18.31 \pm 41.03$ & $1.19 \pm 0.27$ & 0.19 & $3.99 \times 10^{-1}$  & 0.472	\\
$L_{X,5}$ $vs.$ $L_{iso}$    &  $ -3.54 \pm 10.00$ & $0.98 \pm 0.13$ & 0.48 & $2.35 \times 10^{-2}$  & 0.329	\\
$L_{X,1}$ $vs.$ $L_{iso}$    &  $ -4.82 \pm 16.80$ & $0.98 \pm 0.17$ & 0.30 & $1.78 \times 10^{-1}$  & 0.383	\\
$L_{X,11}$ $vs.$ $L_{iso}$   &  $-13.60 \pm 30.76$ & $1.12 \pm 0.23$ & 0.26 & $2.43 \times 10^{-1}$  & 0.461	\\
$L_{X,24}$ $vs.$ $L_{iso}$   &  $-15.72 \pm 36.14$ & $1.15 \pm 0.25$ & 0.22 & $3.20 \times 10^{-1}$  & 0.485	\\
$L_{X,5}$ $vs.$ $E_{peak}$   &  $44.34 \pm 12.53$ &  $1.46 \pm 0.36$ & 0.38 & $7.78 \times 10^{-2}$  & 0.299	\\
$L_{X,1}$ $vs.$ $E_{peak}$   &  $43.71 \pm 10.27$ &  $1.26 \pm 0.39$ & 0.19 & $3.94 \times 10^{-1}$  & 0.342	\\
$L_{X,11}$ $vs.$ $E_{peak}$  &  $43.08 \pm 65.28$ &  $0.97 \pm 0.53$ & 0.06 & $8.06 \times 10^{-1}$  & 0.484	\\
$L_{X,24}$ $vs.$ $E_{peak}$  &  $42.87 \pm 21.20$ &  $0.87 \pm 0.56$ &-0.06 & $8.01 \times 10^{-1}$  & 0.537	\\ \hline
\hline
\end{tabular}
\label{tab:log_corr}
\end{table*}

\section*{Acknowledgments}
We thank the referee, Dr. Bruce Gendre, for his constructive and useful comments and suggestions. This work made use of data supplied by the UK Swift Science Data Centre at the University of Leicester.
We acknowledge the Italian Space Agency (ASI) for financial support through the grant I/011/07/0.

\label{lastpage}


\begin{thebibliography}{99}
\bibitem[]{} Amati, L. et al., 2002, A\&A 390, 81
\bibitem[]{} Berger, E. et al., 2003, ApJ, 590, 379
\bibitem[]{} Berger, E. 2007, ApJ, 670, 1254
\bibitem[]{} Bernardini, M. G. et al., 2012a, A\&A, 539, A3
\bibitem[]{} Bernardini, M. G. et al., 2012b, arXiv:1203.1060
\bibitem[]{} Blundell, K. M., Rawlings, S., Willot, C., 1999, AJ, 117, 677
\bibitem[]{} Bremer, M. et al. 1998, A\&A, 332, L13
\bibitem[]{} Chevalier, R. A., Li, Z., 1999, ApJ, 520, L29
\bibitem[]{} Chevalier, R. A., Li, Z., 2000, ApJ, 536, 195
\bibitem[]{} Costa E. et al. 1997, Nature, 387, 783
\bibitem[]{} Evans, P. et al. 2009, MNRAS, 397, 1177
\bibitem[]{} Frail, D. A. et al. 1997, Nature 389, 261
\bibitem[]{} Freedman, D. L., Waxman, E., 2001, ApJ, 547, 922
\bibitem[]{} Gehrels, N. et al., 2004, ApJ, 611, 1005
\bibitem[]{} Gehrels, N. et al., 2008, ApJ, 689, 1161
\bibitem[]{} Gendre, B. et al., 2008, ApJ, 683, 620
\bibitem[]{} Ghirlanda, G. Ghisellini G., Lazzati D., 2004, ApJ, 616, 331
\bibitem[]{} Ghirlanda, G. et al. 2009, A\&A, 496, 585
\bibitem[]{} Ghirlanda, G. et al. 2012, MNRAS, 420, 483
\bibitem[]{} Granot, J., Konigl, A., Piran, T., 2006, MNRAS, 370, 1946
\bibitem[]{} Heng, K. et al. 2008, ApJ, 681, 1116
\bibitem[]{} Isobe, T. et al., 1990, ApJ, 364, 104
\bibitem[]{} Jakobsson, P. et al. 2006, A\&A, 447, 897
\bibitem[]{} Kann, D. A., et al. 2010, ApJ, 720, 1513
\bibitem[]{} Kann, D. A., et al. 2011, ApJ, 734, 96
\bibitem[]{} Kendall, M., Stuart, A., 1979, The Advanced Theory of Statistics. MacMillian, New York
\bibitem[]{} Kouveliotou, C. et al., 1993, ApJ, 413, L101
\bibitem[]{} Kumar, P. and Panaitescu, A. 2000, ApJ, 541, L51
\bibitem[]{} Kumar, P., 2000, ApJ, 538, L125
\bibitem[]{} Mangano, V. et al., 2007, A\&A, 470, 105
\bibitem[]{} Margutti, R. et al., 2012, arXiv:1203.1059
\bibitem[]{} M\'esz\'aros, P. 2002, ARA\&A, 40, 137
\bibitem[]{} M\'esz\'aros, P. 2006, Reports on Progress in Physics, 69, 2259
\bibitem[]{} Nakar, E., 2007, Phys. Rev., 442, 166
\bibitem[]{} Nava, L. et al. 2012, MNRAS in press, arXiv:1112.4470
\bibitem[]{} Nousek, J. A. et al. 2006, ApJ, 642, 389
\bibitem[]{} Nysewander, M. et al. 2009, 701, 824
\bibitem[]{} O'Brien, P. T. et al. 2006, ApJ, 647, 1213
\bibitem[]{} Padovani, P. 1992, A\&A, 256, 399
\bibitem[]{} Panaitescu, A. and Kumar, P. 2000, ApJ, 543, 66
\bibitem[]{} Press, W. H., Flannery, B. P., Teukolsky, S. A., \& Vetterling, W. T., 1986, Numerical Recipes (New York: Cambridge Univ. Press)
\bibitem[]{} Racusin, J. L. et al. 2011, ApJ, 738, 138
\bibitem[]{} Salvaterra, R. et al. 2012, ApJ in press, arXiv:1112.1700
\bibitem[]{} Sari, R. et al., 1998, ApJ, 497, L17
\bibitem[]{} Sari, R. and Piran, T. 1999, ApJ, 520, 641
\bibitem[]{} Spearman, C. 1904, Am. J. Psycol., 15, 72
\bibitem[]{} Tagliaferri, G. et al., 2005, Nature, 436, 985
\bibitem[]{} van Paradijs, J. et al. 1997, Nature 386, 686
\bibitem[]{} Yonetoku, D. et al., 2004, ApJ, 609, 935
\bibitem[]{} Willingale, R. et al. 2007, ApJ, 662, 1093
\bibitem[]{} Zhang, B. et al. 2006, ApJ, 642, 354

\end{thebibliography}
\end{document}